\documentclass[prl,twocolumn,showpacs, superscriptaddress]{revtex4-1}
\usepackage{placeins}
\usepackage{amsfonts}
\usepackage{amssymb}
\usepackage{amsmath, mathtools}
\usepackage{graphicx} 
\usepackage{datetime}
\usepackage{xcolor}
\usepackage{enumerate}
\usepackage{pifont,array}

\newcommand{\ket}[1]{|#1\rangle}
\newcommand{\bra}[1]{\langle#1|}

\newcommand{\expectn}[1]{\langle#1\rangle}

\begin{document}

\title{The Helstrom measurement: A nondestructive implementation}

\author{Rui {Han}}
\affiliation{Max Planck Institute for the Science of Light, 91058 Erlangen, Germany}
\affiliation{Institute of Optics, Information and Photonics, University of Erlangen-N\"urnberg, 91058 Erlangen, Germany}

\author{Gerd {Leuchs}}
\affiliation{Max Planck Institute for the Science of Light, 91058 Erlangen, Germany}
\affiliation{Institute of Optics, Information and Photonics, University of Erlangen-N\"urnberg, 91058 Erlangen, Germany}
\affiliation{University of Ottawa, Ottawa ON K1N 6N5, Canada}

\author{J\'anos A. {Bergou}}
\affiliation{Department of Physics and Astronomy, Hunter College and the Graduate Center of the City University of New York, 10065 New York, NY, USA}

\begin{abstract}

We discuss a novel implementation of the minimum error state discrimination measurement, originally introduced by Helstrom~\cite{Helstrom:76}. In this implementation, instead of performing the optimal projective measurement directly on the system, it is first entangled to an ancillary system and the measurement is performed on the ancilla. We show that, by an appropriate choice of the entanglement transformation, the Helstrom bound can be attained. The advantage of this approach is twofold. First, it provides a novel implementation when the optimal projective measurement cannot be directly performed. For example, in the case of continuous variable states (binary and N phase-shifted coherent signals), the available detection methods, photon counting and homodyning, are insufficient to perform the required cat-state projection. In the case of symmetric states, the square-root measurement is optimal, but it is not easy to perform directly for more than two states. Our approach provides a feasible alternative in both cases. Second, the measurement is non-destructive from the point of view of the original system and one has a certain amount of freedom in designing the post-measurement state, which can then be processed further.

\end{abstract}


\maketitle

The Helstrom bound is one of the first rigorous results in quantum information theory. It provides the optimal solution to the following problem~\cite{Helstrom:76, HOLEVO1973337,1055351}. Alice randomly prepares a quantum system in one of two states, $\ket{\psi_1}$ and $\ket{\psi_2}$, and sends the system to Bob. The states and their \emph{a priori} probabilities (or priors, in short), $\eta_1$ and $\eta_2$ (such that $\eta_1+\eta_2=1$), are also known to Bob, so he receives $\ket{\psi_i}$ with probability $\eta_i$. Bob's task is to guess, the best he can, the state of the system every time he receives one, possibly aided by a measurement he can perform on the system. This cannot be accomplished without error if the states are not mutually orthogonal and the task is to find the measurement that will identify the state with the smallest error allowed by the laws of quantum mechanics. The optimal error probability for minimum error state discrimination strategy (MESD) is given by the \emph{Helstrom bound}~\cite{Helstrom:76, Helstrom:79},
\begin{eqnarray}
P_\mathrm{E}=\frac{1}{2}\left(1-\sqrt{1-4 \eta_1 \eta_2|\expectn{\psi_1|\psi_2}|^2}\right) .
\label{Helstrombound}
\end{eqnarray}

The impossibility to perfectly discriminate nonorthogonal states is central to quantum communication schemes and, in particular, to security of quantum key distribution.
Various strategies have been proposed for discrimination of nonorthogonal states~\cite{Barnett:09, Bergou:10, Bae:15,Weir2017}. The existing methods are remarkably successful for many state discrimination problems with discrete variables. However, there are limitations that are difficult to overcome in discriminating continuous variable states. The limitations come from both the available detectors and their efficiencies.

When an observer performs a standard projective measurement (rank-1 projector) on a system, the state of the system often `collapses' at the detector. The so-called post-measurement state does not only becomes an eigenstate of the projector but also it is often completely destroyed by the detector such that no residual states escapes from the detector. The measurement is, thus, destructive, and it is generally assumed that any information about the state before the measurement is lost in the process (for an alternate view, however, see \cite{rapcan}).

The purpose of this paper is to show that this commonly accepted view of standard quantum measurements can be significantly refined. We present an alternative derivation of the Helstrom bound based on a non-destructive implementation of positive-operator-valued-measurements (POVMs). In this implementation, Bob first entangles the system with an ancilla qunit in such a way that the information carried by the system is transferred to the ancilla, and a projective measurement is then performed on the (discrete) ancilla. 

In addition to offering a simple mathematical derivation of the Helstrom bound, this implementation also yields two significant advances. First, it offers a solution to the discrimination of systems for which the direct physical implementation of the Helstrom measurement is not available. For example, when discriminating continuous variable states, such as coherent states, it is notoriously difficult or outright impossible to reach the Helstrom bound. The optimum measurement would require projections to `cat states' but the only available detections, photon counting and homodyning, are insufficient to implement the required projections~\cite{Kennedy1973, Dolinar1973, Bondurant:93}. Based on the currently available detection techniques, bounds less tight than Eq. \eqref{Helstrombound} were established and some of them were demonstrated in recent experiments~\cite{Cook2007, PhysRevLett.104.100505, PhysRevLett.106.250503, 1367-2630-14-8-083009, Becerra14}. Second, since the measurement is on the ancilla and, thus, nondestructive for the system, there is a certain amount of information left in the post-measurement state if the measurement is not optimal. Therefore, it offers the flexibility of implementing sequential measurements~\cite{PhysRevLett.111.100501}.

In this paper, we first show how to construct the nondestructive implementation for the discrimination of two pure quantum states, and that it can saturate the Helstrom bound. Next, we extend the scheme to the discrimination of two other classes of quantum states for which the lower bound and the Helstrom measurements are theoretically known, namely, the discrimination of $N$ real symmetric states (definitions see below) and the discrimination of $N$ phase-shifted coherent states. For these two classes, we show that the lower bound of the error probability, given by the optimal measurement operators~\cite{Ban1997, PhysRevA.54.1691}, can be attained with the nondestructive scheme. The scheme not only gives alternative mathematical constructions of the minimum error probabilities, it also provides an alternative approach for the physical implementations using ancilla systems. We conclude with a summary and outlook.


The nondestructive approach employs the Neumark extension~\cite{Neumark}, which has been routinely used in the implementation of POVMs, for the implementation of the optimal measurement (some of the ideas of the present paper were already introduced in~\cite{PhysRevLett.111.100501}, where a theory of sequential quantum measurements has been developed). In this approach, instead of performing the measurement directly on the state sent by Alice, Bob first prepares an ancilla in some initial state $\ket{i}$ and applies a unitary transformation that entangles the state he received with the ancilla. In the general scenario, Alice encodes her message using a set of $N$ pure quantum states $\{\ket{\psi_j}$, $j=1,2,\dots,N\}$, with prior probabilities $\{\eta_j,j=1,2,\dots,N\}$, the unitary transformation between the signal and ancilla is
\begin{equation}
U\ket{\psi_j}\ket{i}=\sum_{k=1}^N c_{jk}\ket{\varphi_{jk}}\ket{k}\;\;\;\mathrm{for}\;\;\;j=1,2,\dots,N\,,\label{eq:Unitary}
\end{equation}
where $\{\ket{k}, k=1,2,\dots,N\}$ forms an orthonormal basis for the ancilla space. The unitary is constructed in such a way that the diagonal amplitudes $c_{jj}$ are as large as possible, i.e., as permitted by the constraints imposed by the laws of quantum mechanics. After the unitary transformation, Bob performs standard projective quantum measurements on the ancilla with the projectors $\{P_j=\ket{j}\bra{j}, j=1,2,\dots,N\}$ and identifies his state with $\ket{\psi_j}$ if $P_j$ clicks. Hence, the probability that Bob identifies the state correctly is $P_\mathrm{succ}=\eta_j\sum_{j=1}^N |c_{jj}|^2$, where $\eta_j$ is the prior probability of state $\ket{\psi_j}$, and the error probability is
\begin{equation}
P_\mathrm{err}=1-\eta_j\sum_{j=1}^N |c_{jj}|^2\,.\label{eq:NsymPerr}
\end{equation}
The task is minimize $P_\mathrm{err}$ with the constraints given by the unitarity of transformation \eqref{eq:Unitary}.

For the discrimination of binary states, the Neumark expension can be described by
\begin{eqnarray}
U\ket{\psi_1}\ket{i}&=&\sqrt{p_1}\ket{\varphi_1}\ket{1}+\sqrt{r_1}\ket{\phi_1}\ket{2}\,,\nonumber\\
U\ket{\psi_2}\ket{i}&=&\sqrt{r_2}\ket{\varphi_2}\ket{1}+\sqrt{p_2}\ket{\phi_2}\ket{2}\,,\label{unitary}
\end{eqnarray}
So, when Bob performs an orthogonal measurement on the ancilla in the $\{\ket{1},\ket{2}\}$ basis, he will guess the input as $\ket{\psi_{1}}$ if the outcome is $\ket{1}$ and $\ket{\psi_{2}}$ if the outcome is $\ket{2}$.  
and, the average probability of error is
\begin{eqnarray}
P_\mathrm{err}= \eta_1r_1+ \eta_2r_2\,.\label{eq:PerrJanos}
\end{eqnarray}

Let us first discuss the role of the post-measurement states. Clearly, if Alice sent $\ket{\psi_1}$ then $\ket{\varphi_1}$ is the state of the system after the measurement if the ancilla is found in $\ket{1}$ and $\ket{\phi_1}$ if the ancilla is found in $\ket{2}$. Similarly, $\ket{\varphi_2}$ and $\ket{\phi_2}$ are the post-measurement states if $\ket{\psi_2}$ was sent. When Bob finds the ancilla in $\ket{1}$, the state of the qubit is either $\ket{\varphi_1}$ or $\ket{\varphi_2}$. If these states were different Bob could perform further discrimination of the post-measurement states, gaining further information on the initial preparation. So, his measurement would not be optimal since discrimination of the post-measurement states of the system would reduce the probability of error. From here it follows that we must require $\ket{\varphi_1}=\ket{\varphi_2}$ and $\ket{\phi_1}=\ket{\phi_2}$, for optimal discrimination.

Since the transformation is unitary we have $p_1+r_1=1$ and $p_2+r_2=1$, so we either correctly identify the state or make an error. By taking the inner product of the two equations in~(\ref{unitary}), we have 
\begin{equation}
s\equiv\expectn{\psi_1|\psi_2}=\sqrt{(1-r_1)r_2}+\sqrt{(1-r_2)r_1}\,,\label{eq:sConstraint}
\end{equation}
which is the constraint for the minimization of $P_\mathrm{err}$. 
Before proceeding to the general solution, we notice that for the case of equal priors, $\eta_1= \eta_2=\frac{1}{2}$, no further optimization is necessary. In this case the problem is symmetric in the two inputs, so we can assume $r_1=r_2=r$ and the above equation immediately yields $s=2\sqrt{(1-r)r}$. Solving this equation for $r$ gives the average error probability $r=\frac{1}{2}(1-\sqrt{1-|\expectn{\psi_1|\psi_2}|^2})=P_\mathrm{E}$.
which is the Helstrom bound for equal priors.

For arbitrary priors we use the method of Langrange multipliers. The quantity to be optimized becomes
\begin{equation}
P_\mathrm{err}^{\lambda}= \eta_1r_1+ \eta_2r_2+\lambda\Big(s-\sqrt{(1-r_1)r_2}-\sqrt{(1-r_2)r_1}\Big).
\end{equation}
We follow the usual procedure by taking the derivative of the above expression with respect to $\lambda$, $r_1$ and $r_2$, respectively, and set them equal to 0. The first one just yields the constraint (\ref{eq:sConstraint}), and the other two yield
\begin{eqnarray*}
 \eta_1\sqrt{r_1(1-r_1)}&=&\frac{\lambda}{2}\left(\sqrt{(1-r_1)(1-r_2)}-\sqrt{r_1r_2}\right),\\
 \eta_2\sqrt{r_2(1-r_2)}&=&\frac{\lambda}{2}\left(\sqrt{(1-r_1)(1-r_2)}-\sqrt{r_1r_2}\right).
\end{eqnarray*}
The LHS of the first equation is independent of $r_2$ and the LHS of the second equation is independent of $r_1$. For the two expressions to be equal (as it is for their RHS), the LHS expressions must be a constant $c$ that is independent of both $r_1$ and $r_2$. Solving it together with the constraint, we have $c^2= \eta_1^2 \eta_2^2s^2(1-s^2)/(1-4 \eta_1 \eta_2s^2)$, and 
\begin{equation}
r_{1,2}=\frac{1}{2}\left(1-\frac{1-2\eta_{2,1}s^2}{\sqrt{1-4 \eta_1 \eta_2s^2}}\right).
\end{equation}
Using these optimized individual error probabilities in Eq.~\eqref{eq:PerrJanos}, immediately yields the Helstrom bound, Eq.~\eqref{Helstrombound}.
It is worth noting that not only the overall error probability but also the individual error probabilities are exactly the same as those of the optimal Helstrom measurement.

Here, we would like to point out the two main advantages of the nondestructive scheme. First, it allows for a simple derivation of the minimum error probability for any set of pure binary signals, especially when the prior probabilities are equal. Second, instead of implementing the Helstrom measurement, which is complicated to construct for some systems, systems with continuous variables for example, it just requires standard orthogonal projective measurements on the ancilla. The complication is shifted to finding the suitable easy-to-measure ancilla qunit and implementing the optimal unitary operation $U$. For example, the signal states can be entangled to discrete atomic ancilla via atom-light interaction, or they can be entangled to different degrees of freedom (or different fields) using non-linear medium. A near optimal discrimination of binary coherent signals via atom-light interaction is demonstrated in Ref.~\cite{RLB2}.


As our next example, we consider the case in which Alice encodes her message using a set of $N$ real symmetric states $\{\ket{\psi_j}$, $j=1,2,\dots,N\}$, with equal priors $\eta_j=1/N$. The term `real symmetric' implies that the overlaps of any two of these states are equal and real,
\begin{equation}
s\equiv\expectn{\psi_j|\psi_k} = s^{*} \in\Re\,,\;\;\;\;\mathrm{for}\;\; j\ne k\,.
\end{equation}
These states are also referred to as the edges of a quantum pyramid where $s$ is the cosine of the angle of the pyramid~\cite{Englert2010}, or equidistance states~\cite{Roa11}. In general, the set of $N$ real symmetric states used by Alice are linearly independent and spans an $N$-dimensional Hilbert space, except in the limiting case $s=-1/(N-1)$ when the dimensionality of the states is reduced to $N-1$ and the states are linearly dependent. 
Other than its applications in quantum cryptography, the real symmetric states also appear in many other applications of quantum information science~\cite{Jimenez10, Paiva10, Sehrawat2011}. Thus, it is not only a system of theoretical interest but also of practical significance.

Such discrimination problems have been intensively studied both analytically and numerically. The optimal measurement that minimizes the average error probability for any set of real symmetric states is given by the well-know square-root measurement (SRM)~\cite{Ban1997}. Although a closed-form analytical expression of SRM is available, the optimal measurements are generally challenging to implement as they require projective measurements onto superpositions of the entire set of $N$ states~\cite{Solis2017}.
Here, we apply the nondestructive measurement scheme to the discrimination of an arbitrary set of real symmetric quantum states and show that the minimum error probability given by SRM can be attained with projective measurements on the ancilla system in a straightforward manner.


Since the set of states to discriminate is real and symmetric, there is no reason to introduce asymmetry for the unitary operation. Furthermore, one can restrict the coefficients of the ancilla state $\ket{i}$ to be real and having only real coefficients $c_{jk}$s. Thus, we can set $c_{jj}=\sqrt{p}$ and $c_{jk}=\sqrt{r}$ for $k\neq j$, where $p+(N-1)r=1$ is required by the unitarity of the process.
Following our arguments of the previous example, optimal discrimination is achieved when the post-measurement states are identical for any given measurement outcome, because no further information is contained in the post-measurement states. Thus, the optimal discrimination requires
\begin{eqnarray}
\begin{array}{c}U\ket{\psi_1}\ket{i}=\sqrt{p}\ket{\varphi_1}\ket{1}+\sqrt{r}\ket{\varphi_2}\ket{2}+\ldots+\sqrt{r}\ket{\varphi_N}\ket{N},\\
U\ket{\psi_2}\ket{i}=\sqrt{r}\ket{\varphi_1}\ket{1}+\sqrt{p}\ket{\varphi_2}\ket{2}+\ldots+\sqrt{r}\ket{\varphi_N}\ket{N},\\[-1ex]
\vdots\\
U\ket{\psi_N}\ket{i}=\sqrt{r}\ket{\varphi_1}\ket{1}+\sqrt{r}\ket{\varphi_2}\ket{2}+\ldots+\sqrt{p}\ket{\varphi_N}\ket{N}.\end{array}\nonumber\\
\label{eq:NsymUopt}
\end{eqnarray}
In this case the error probability is $P_\mathrm{err}=1-p$, and the task is to maximize $p$.

\begin{figure}[h]
\centerline{\setlength{\unitlength}{1pt}
\begin{picture}(220,135)(0,0)
\put(5,-5){\includegraphics[scale=0.52]{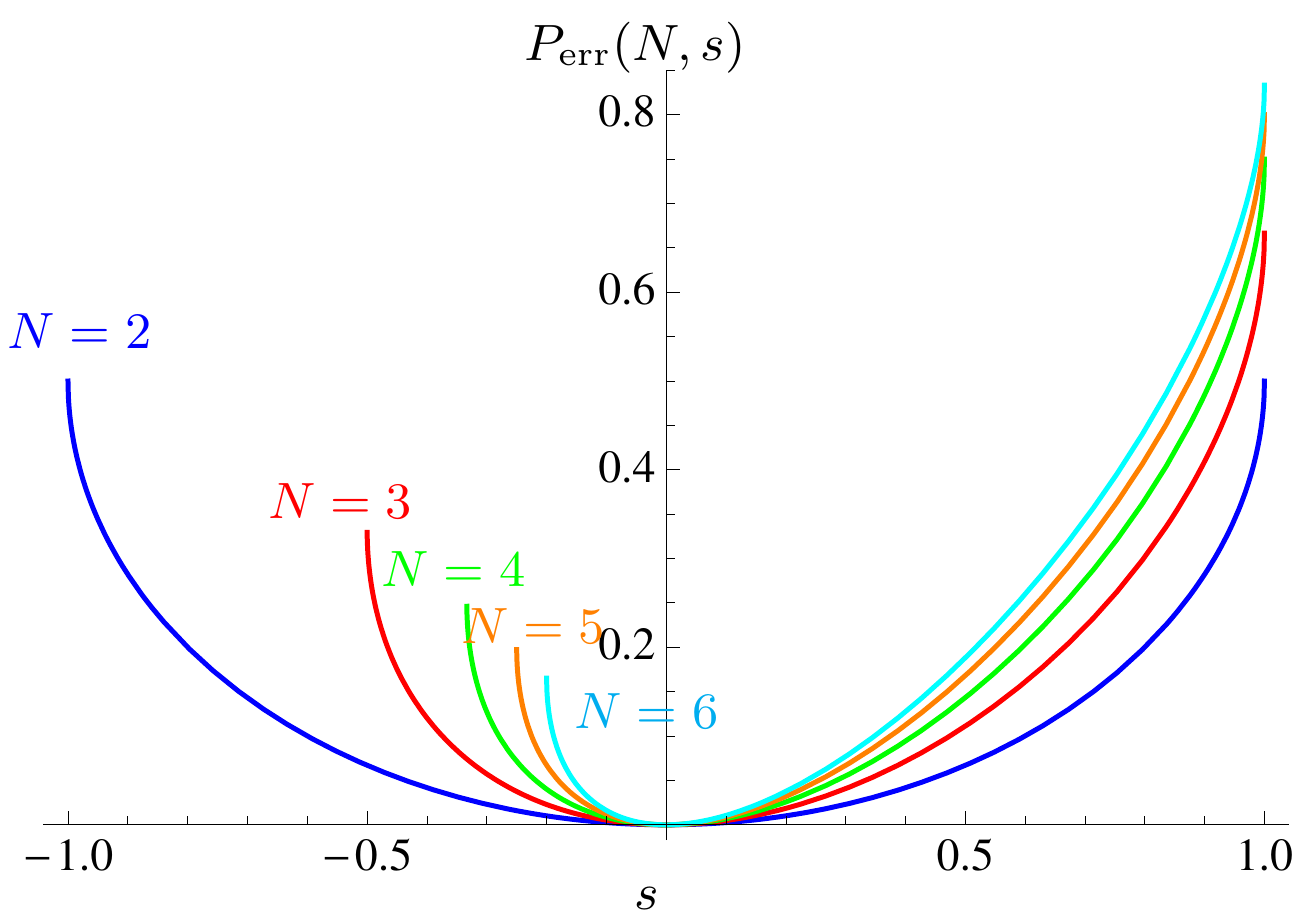}}
\end{picture}}
\caption{The minimum error probability $P_\mathrm{err}^{\mathrm{min}}(N,s)$ for discrimination among $N$ real symmetric states for $N=2,3,4,5,6$ as a function of the overlap between the states, $s\equiv\expectn{\psi_j|\psi_k}$ (for $j\neq k$).}\label{fig:NsymPerrNs}
\end{figure}

In general, a set of $N(N-1)/2$ equations obtained from taking pairwise inner product of the $N$ equations in~(\ref{eq:Unitary}) serves as the constraints to minimize $P_\mathrm{err}$, which could make the optimization problem highly nontrivial. 
Fortunately, for the unitary process described by Eqs.~(\ref{eq:NsymUopt}), taking the inner product of any two equations gives the same constraint, i.e.,
$s=2\sqrt{pr}+(N-2)r$.
This yields a quadratic equation in $p$ to solve,
and the solutions are
\begin{equation}
p=\frac{1}{N^2}\left[\sqrt{1+s(N-1)}\pm(N-1)\sqrt{1-s}\right]^2.
\end{equation}
Both solutions are positive and the larger solution among the two, which is the one with the `$+$' sign, gives the minimum error probability
\begin{equation}
P_\mathrm{err}^{\mathrm{min}}(N,s)=1-\frac{1}{N^2}\left[\sqrt{1{+}s(N{-}1)}+(N{-}1)\sqrt{1{-}s}\right]^2.\label{eq:NsymPerrNs}
\end{equation}
This simple analytical expression of $P_\mathrm{err}^{\mathrm{min}}$ for discrimination among any set of $N$ real symmetric states with transition amplitude $s$ among the states is illustrated by Fig.~\ref{fig:NsymPerrNs}. As expected, it agrees with the expression obtained from SRM~\cite{Ban1997,Englert2010, Herzog2012} that requires the implementation of projective measurements $P_j=\ket{\mu_j}\bra{\mu_j}$, where
$\ket{\mu_j}=\Phi^{-1/2}\ket{\psi_j}\;\;\;\mathrm{with}\;\;\;\;\Phi=\sum_{j=1}^N\ket{\psi_j}\bra{\psi_j}$. 

As our final example, we show that our scheme also provides an easy implementation for the discrimination of coherent-state signals obtained from phase shift keying -- another class of widely used systems for classical and quantum communications~\cite{PhysRevLett.88.057902, Grosshans03} and implementations of quantum information science~\cite{RevModPhys.84.621, PhysRevLett.93.250502, 1367-2630-7-1-137, LPOR:LPOR201000005, Becerra2013, Muller2015}. 
Suppose Alice encodes her information using a set of $N$ coherent states $\{\ket{\psi_j}=\ket{e^{2\pi i\frac{j-1}{N}}\alpha},j=1,2,\dots,N\}$ with equal priors.
These coherent states have the same intensity $|\alpha|^2$ and the information is encoded in the symmetrically distributed phases. The binary case covered by our first example has been well-studied analytically and the minimum error probability is given by the Helstrom bound. For information encoded in a set of more than two of such states, the minimum error probability is obtained with the optimum detection operators given by SRM. The explicit analytical solutions for ternary and quaternary phase shifted signals have been derived in~\cite{PhysRevA.54.1691}.

Following the procedure of the nondestructive measurement scheme, Bob entangles the signal state with the ancilla qunit with the unitary described in Eq.~(\ref{eq:Unitary}). The optimal unitary should preserve the symmetry of the set of phase-shifted coherent states. In the case of $N=3$, discrimination among $\{\ket{\alpha},\ket{e^{2\pi i/3}\alpha},\ket{e^{-2\pi i/3}\alpha}\}$, the symmetry suggests
\begin{equation}
U\!\left[\!\begin{array}{c}\ket{\psi_1}\\\ket{\psi_2}\\\ket{\psi_3}\end{array}\!\right]\!\ket{i}\!=\!\!\left[\!\begin{array}{ccc}\sqrt{p} & \sqrt{r}e^{i\theta} & \sqrt{r}e^{-i\theta}\\ \sqrt{r}e^{-i\theta} & \sqrt{p} & \sqrt{r}e^{i\theta} \\  \sqrt{r}e^{i\theta} & \sqrt{r}e^{-i\theta} & \sqrt{p}\end{array}\!\right]\!
\!\left[\!\begin{array}{c}\ket{\varphi_1}\ket{1}\\\ket{\varphi_2}\ket{2}\\\ket{\varphi_3}\ket{3}\end{array}\!\!\right]\!.\label{eq:unitary4}
\end{equation}
for real probabilities $p$, $r=(1-p)/2$ and angle $\theta$. The average error probability $P_\mathrm{err}=1-p$ can be minimized with the constraint given by the pairwise inner product of the equations above,
\begin{equation}
s=e^{-\frac{3}{2}|\alpha|^2}e^{i\frac{\sqrt{3}}{2}|\alpha|^2}=2\sqrt{pr} e^{-i\theta}+r e^{2i\theta}\,.
\end{equation}
We now have an optimization problem of two real parameters $p$ and $\theta$ with two real constraints given by the real and imaginary parts of the complex equation above. Let $r_1=\sqrt{r}\cos\theta$ and $r_2=\sqrt{r}\sin\theta$, we have
\begin{eqnarray}
\sqrt{p}+2(r_1+r_2)&=&1\,,\nonumber\\
2\sqrt{p}r_1+r_1^2-r_2^2&=&e^{-\frac{3}{2}|\alpha|^2}\cos{(\sqrt{3}|\alpha|^2/2)}\,,\\
-2\sqrt{p}r_2+2r_1r_2&=&e^{-\frac{3}{2}|\alpha|^2}\sin{(\sqrt{3}|\alpha|^2/2)}\,.\nonumber
\end{eqnarray}
This is exactly the set of constraints for SRM. The analytical solutions are explicitly given in Ref.~\cite{PhysRevA.54.1691}.

\begin{figure}[h]
\centerline{\setlength{\unitlength}{1pt}
\begin{picture}(230,132)(0,0)
\put(0,0){\includegraphics[scale=0.5]{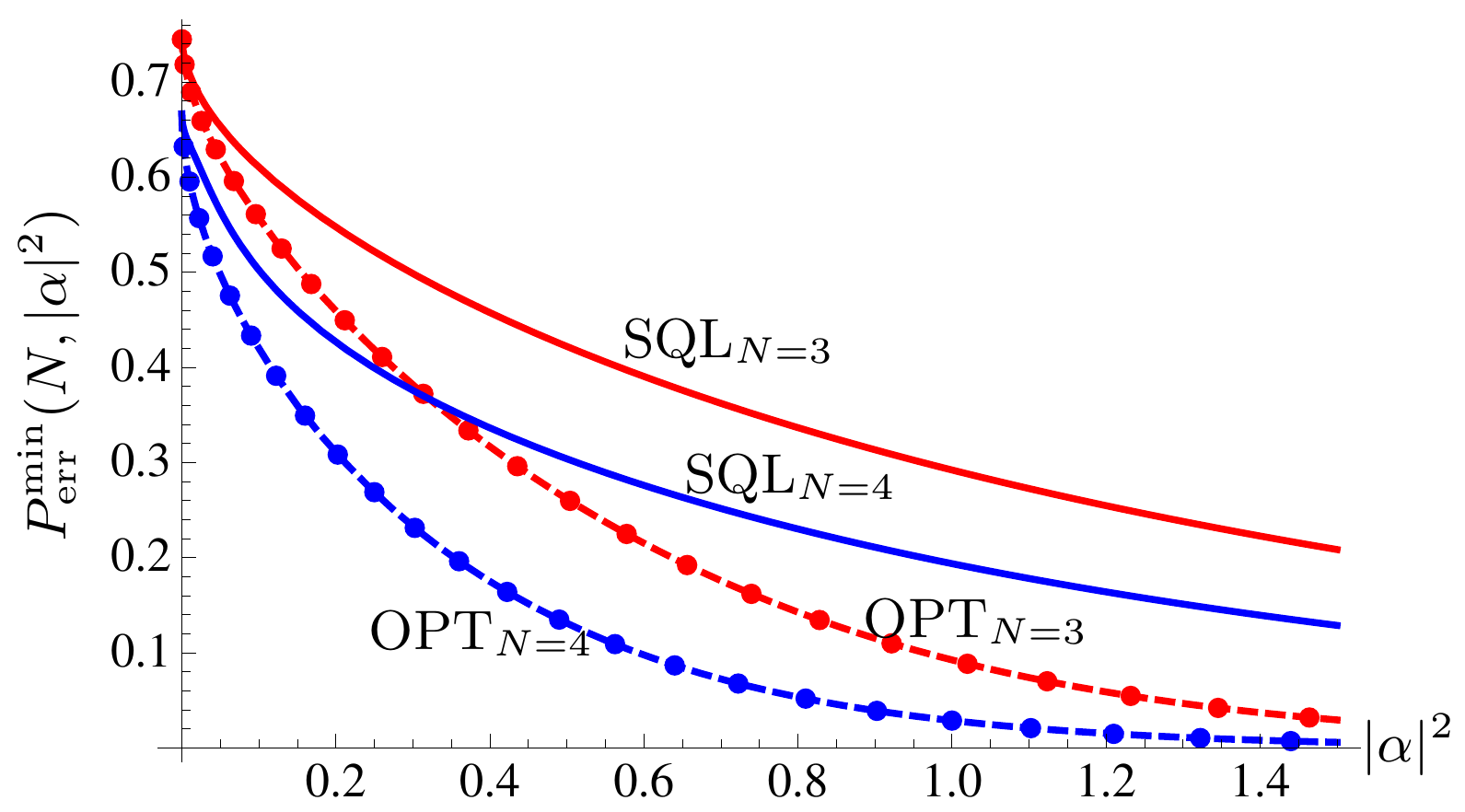}}
\end{picture}}
\caption{The error probability $P_\mathrm{err}^{\mathrm{min}}(N,|\alpha|^2)$ vs. $|\alpha|^2$ for discrimination among $N$ phase-shifted coherent states. The solid curves are the standard quantum limits achievable with perfect homodyne detectors, the dashed curves show the minimum error probability given by the optimum operator measurements~\cite{PhysRevA.54.1691}, and the dots indicate the minimum error probabilities given by the present optimum nondestructive measurement scheme, which are the same as those given by the dashed curves.}\label{fig:CoherentSDN3N4}
\end{figure}

In the case of quaternary signal set $\{\ket{\alpha},\ket{i\alpha},\ket{-\alpha},\ket{-i\alpha}\}$, the pairwise inner products of the signal states are $\expectn{\beta|-\beta}=e^{-2|\beta|^2}$ and $\expectn{\beta|\pm i\beta}=e^{-|\beta|^2(1\pm i)}$, for $\beta=\{\alpha, i\alpha, -\alpha, -i\alpha\}$. This pairwise symmetry of the signal set suggests that the optimal unitary coupling should have coefficients $c_{12}=c_{14}^*=\sqrt{r}{e^{i\theta_1}}$ and independent coefficient $c_{13}=\sqrt{r'}e^{i\theta_2}$. Thus, the optimal unitary operation, suggested by the permutation symmetry of the signal set, is of the form
\setlength{\arraycolsep}{1.0pt}
\begin{equation}
U\!\!\left[\!\begin{array}{c}\ket{\psi_1}\\\ket{\psi_2}\\\ket{\psi_3}\\\ket{\psi_4}\end{array}\!\right]\!\!\ket{i}\!=\!\!\left[\!\!\begin{array}{cccc}\sqrt{p} & \sqrt{r}e^{i\theta_1} & \sqrt{r'}e^{i\theta_2} & \sqrt{r}e^{-i\theta_1}\\ \sqrt{r}e^{-i\theta_1} & \sqrt{p} & \sqrt{r}e^{i\theta_1} & \sqrt{r'}e^{i\theta_2}\\ \sqrt{r'}e^{i\theta_2} & \sqrt{r}e^{-i\theta_1} & \sqrt{p} & \sqrt{r}e^{i\theta_1}  \\ \sqrt{r}e^{i\theta_1} & \sqrt{r'}e^{i\theta_2} &  \sqrt{r}e^{-i\theta_1} & \sqrt{p}\end{array}\!\right]\!
\!\!\left[\!\begin{array}{c}\ket{\varphi_1}\ket{1}\\\ket{\varphi_2}\ket{2}\\\ket{\varphi_3}\ket{3}\\\ket{\varphi_4}\ket{4}\end{array}\!\!\right]\!\!.\label{eq:unitary4}
\end{equation}
These yield the same set of equations to solve for the optimization of $P_\mathrm{err}$ (with real parameters $p$, $r$, $r'$, $\theta_1$, and $\theta_2$) as for the SRM, which can be solved analytically with five real constraints given by Eq.~(\ref{eq:unitary4}). The optimized minimum error probabilities for ternary and quaternary phase-shifted coherent signals are shown in Fig.~\ref{fig:CoherentSDN3N4}.
\setlength{\arraycolsep}{3.0pt}

For simplicity, only examples with equal priors were explicitly discussed above. The scheme, however, can be extended to the discrimination of $N$ states with arbitrary priors. 
The average error probability $P_\mathrm{err}=1-\sum_{j=1}^N \eta_{j}|c_{jj}|^2$ needs to be minimized under the $N(N-1)/2$ constraints that result by taking 
the pairwise inner products of the $N$ equations in Eqs.~\eqref{eq:Unitary}. The coefficients $c_{jk}$ can, in principle, be obtained in the same way as as before, although often only numerically.


In summary, we propose a nondestructive implementation of the Helstrom measurement which is optimal to discriminate two pure quantum states with minimum error. We also demonstrate that the method can be extended to the implementation of SRMs for the discrimination between any set of real symmetric quantum states or any set of phase-shifted coherent states with equal prior probabilities.
More importantly, it is shown that, instead of constructing the complicated and destructive SRMs on the signal states directly, the optimal measurements can be implemented with simple projective measurements on the ancilla and they are nondestructive from the point of view of the system. The challenge is shifted from the construction of optimum measurement operators on the signal to the construction of optimal interaction between the system and ancilla.

\begin{acknowledgments}
       \emph{Acknowledgments}. The research presented in this paper was supported by a grant from the John Templeton Foundation. 
Partial financial support was provided by a grant from PSC-CUNY. Support for a sabbatical stay of JB at the Max Planck Institute for the Physics of Light, Erlangen, Germany, where this research was initiated, is also gratefully acknowledged. 
\end{acknowledgments}

\end{document}